\documentstyle[11pt,aaspp4]{article}

\begin{document}

\title{Constraining the Beaming of $\gamma$-Ray Bursts With Radio Surveys}
\author{Rosalba Perna and Abraham Loeb}
\medskip
\affil{Harvard-Smithsonian Center for Astrophysics, 60 Garden Street,
Cambridge, MA 02138}
%\altaffiltext{2}{email:aloeb@cfa.harvard.edu}

\begin{abstract}

The degree of beaming in Gamma-Ray Bursts (GRBs) is currently unknown.  The
uncertainty in the $\gamma$-ray beaming angle, $\theta_{\rm b}$, leaves the
total energy release ($\propto \theta_{\rm b}^2$) and the event rate per
galaxy ($\propto \theta_{\rm b}^{-2}$) unknown to within orders of
magnitude. Since the delayed radio emission of GRB sources originates from
a mildly relativistic shock and receives only weak relativistic beaming,
the rate of radio-selected transients with no GRB counterparts can be used
to set an upper limit on $\theta_{\rm b}^{-2}$.  We find that a VLA survey
with a sensitivity of $\sim 0.1$ mJy at 10 GHz could identify $\ga 2\times
10^4 (\theta_{\rm b}/10^\circ)^{-2}$ radio afterglows across the sky if
each source is sampled at least twice over a period of one month or longer.
From the total number of $\ga$ 0.1 mJy sources observed at 8.44 GHz and the
fraction of fading sources at 1.44 GHz, we get the crude limit $\theta_{\rm
b}\ga 6^\circ$.

\end{abstract} 


\section{Introduction}

Gamma-Ray Burst (GRB) sources were discovered historically in $\gamma$-rays
because they are rare and hence require continuous monitoring of large
areas of the sky. The necessary all-sky monitoring program was first made
feasible in the $\gamma$-ray regime, hence the name GRBs.  By now, the
prompt $\gamma$-ray emission is known to be followed in most cases by
delayed emission in the $X$-ray (Costa et al.  1997), optical (van Paradijs
et al. 1997; Bond 1997a,b; Fruchter et al.  1998a) and radio (Frail et al.
1997a,b; Frail et al. 1998a,b; Kulkarni et al. 1998b; Taylor at al. 1998)
bands.  Given the recent discovery of afterglows, it is now timely to
explore the feasibility of searching for GRB sources at longer wavelengths
than traditionally attempted.  The importance of complementary searches is
highlighted by the possibility that some sources might be $\gamma$-ray
faint for geometric or physical reasons. Such sources can only be found
at long wavelengths, e.g. in radio surveys.

All past discoveries of GRB afterglows were selected by the detection of
$\gamma$-rays. If the $\gamma$-ray emission is beamed, then there should be
a population of radio afterglows which are $\gamma$-ray faint (Rhoads
1997).  This follows from the fact that the bulk Lorentz factor of the
emitting material in GRB sources declines with time and is only of order
unity when the late radio emission takes place (Waxman, Kulkarni, \& Frail
1998), hence making the effect of relativistic beaming weak at that time.
In popular synchrotron models (e.g. Meszaros \& Rees 1997; Waxman 1997a,
1997b), the gradual drift of the peak flux to lower frequencies as a
function of time is caused by the decline of the Lorentz factor of the
emitting material, and so the minimum solid angle into which the radiation
could be focused increases with time.  If the $\gamma$-ray emission is
beamed into a fraction $f_{\rm b}$ of the sky, then the rate of radio
afterglows could be larger than the measured rate of GRBs by up to $\sim
f_{\rm b}^{-1}$.  For relativistic expansion with a Lorentz factor
$\Gamma$, the radiation observed at a given time $t$ originates from a
conical section of opening angle $\sim 1/\Gamma(t)$ of the fireball. The
smallest value possible for $f_{\rm b}$ is therefore $\sim \pi
\Gamma^{-2}/4\pi=1/4\Gamma^{2}$, where $\Gamma$ is the Lorentz factor at
the time of the $\gamma$-ray emission.  Since typical GRB sources have
$\Gamma\ga 10^2$ (e.g., Fenimore, Epstein, \& Ho 1993; Woods \& Loeb 1995)
while the radio afterglow emission occurs when $\Gamma \sim 1$, the rate of
radio afterglows could be boosted by $\ga 4$ order of magnitudes relative
to the measured GRB rate.  The total energy associated with a GRB explosion
is proportional to $f_{\rm b}$ and is similarly uncertain. Thus, the
determination of $f_{\rm b}$ is of crucial importance for identifying the
likely astrophysical origin of GRBs.

The evolution of a jet resembles that of a spherical fireball (i.e. the jet
behaves as a conical section of a spherical fireball), as long as the
expansion Lorentz factor is larger than the inverse of the jet opening
angle, $\Gamma>\theta_{\rm b}^{-1}$.  The smooth power-law decline of
optical afterglows, observed over a time scale of days to months for
GRB970508 and GRB970228 (Livio et al. 1997; Sokolov et al. 1998), suggests
that these fireballs behaved as if they were spherically symmetric on
angular scales ranging from $1/\Gamma(1{\rm day})\sim 1/8$ to
$1/\Gamma\;(1\;{\rm month})\sim1/3$ (Rhoads 1997, 1998; Waxman 1997a).
However, there is still a missing gap in afterglow observations on the time
window of minutes to several hours following a GRB. During this time the
fireball decelerates from $\Gamma\sim100$ to $\Gamma\sim10$, and so there
is uncertainty about the structure of the fireball on angular scales
$\sim0.01$--0.1 radians.  It is still possible that the highly relativistic
expansion with $\Gamma\sim10^2$ is restricted only to a very small angular
diameter, $\theta_{\rm b}\sim 1/\Gamma=0.6^{\circ}$.  In fact, the popular
GRB scenarios of binary coalescence of compact stars or failed supernovae
favor strong collimation over spherical expansion (Woosley 1998; Fryer et
al. 1998; Meszaros et al. 1998). The search for long-wavelength transients
with no GRB counterparts can set important constraints on $\theta_{\rm b}$
and hence on these models.

Current long-wavelength surveys only place weak limits on the beaming
factor.  In particular, existing supernova searches imply that at most
$\sim 10\%$ of all transient optical events cannot be classified as having
classical supernovae lightcurves (Kirshner 1998, private communication).
Since the supernova rate per galaxy is $\sim 4$ orders of magnitude higher
than the GRB rate (Wijers et al. 1997), the enhancement factor due to
beaming in the optical band must be $\la 10^3$ (see Woods \& Loeb 1998 for
a more detailed discussion). The enhancement factor could still be an order
of magnitude higher in the radio than in the optical.  Previous radio
experiments such as CLFRT at 0.151 GHz (Dessenne et al. 1996), and COBE at
90 GHz (Ali et al. 1997) do not provide useful limits on this factor, as
they achieved only a sensitivity of 10--100 Jy which is well above the
characteristic mJy radio flux of GRB afterglows.

In this {\it Letter}, we show that a $0.1$ mJy radio survey at the VLA
[analogous to the FIRST survey (Helfand et al. 1996)] can provide strong
constraints on $\theta_{\rm b}$.  The surveyed sources need to be sampled
at least twice over a timescale of order a month or longer, since short
term variability may also be caused by scintillations of steady
pointlike sources.

In our calculations, we assume that the GRB rate is proportional to the
cosmic star formation rate and neglect the small fraction ($\la 10\%$)
of GRBs which might be related to local radio supernovae (Bloom et
al. 1998).  We model the time-dependent luminosity of the radio afterglows
based on observational data and allow for a scatter in their peak
luminosities. Our model is described in detail in \S 2. The number counts
of radio afterglows that it predicts for radio surveys are calculated in \S
3. Finally, \S 4 summarizes our main conclusions.

\section{Statistics of Radio GRB Afterglows} 

The isotropy of GRBs (Meegan et al. 1993; Briggs at al.  1993) and the
flattening of their number count distribution at faint fluxes suggest that
most GRBs occur at cosmological distances. This hypothesis has been
confirmed by the detection of Fe II and Mg II absorption lines at a
redshift of $z=0.835$ in the optical spectrum of GRB 970508 (Metzger et
al. 1997), the inference of a redshift $z=3.42$ for the host galaxy of GRB
971214 (Kulkarni et al. 1998a), and the detection of emission and
absorption lines at $z=0.966$ for the host of GRB980703 (Djorgovski et
al. 1998).

The energy scale of cosmological GRBs, $\sim 10^{53} f_{\rm b}~{\rm ergs}$
(Kulkarni et al. 1998a), corresponds to the rest mass energy of a fraction
of a solar mass, and implies a link between their energy budget and compact
stars.  Indeed, the most popular models for the origin of GRBs relate them
to compact stellar remnants, such as neutron stars or black holes (e.g.
Paczy\'nski 1986, 1998; Eichler et al.  1989; Narayan et al. 1992; Usov
1992; Woosley 1993; Rees 1998; Popham, Woosley, \& Fryer 1998; Meszaros et
al. 1998) .  Since these remnants form out of massive stars not long after
their birth, it is reasonable to assume that the GRB rate traces the star
formation rate without a significant delay.  This scenario was investigated
by Wijers et al.  (1997), who derived a best-fit constant of
proportionality between the GRB occurrence rate, $R(z)$, and the star
formation rate per comoving volume, $\rho(z)$, based on the requirement
that the former fits the observed number count distribution of GRBs. The
cosmic star formation rate as a function of redshift has been calibrated
based on observations of the $U$ and $B$-band luminosity density evolution
in the Hubble Deep Field (Madau et al.  1996; Madau 1996; Madau 1997;
Madau, Pozzetti, \& Dickinson 1998). The fit made by Wijers et al. (1997)
yields a local GRB rate of $R(z=0)= (0.14\pm 0.02) \times 10^{-9} ~{\rm
Mpc}^{-3} {\rm yr}^{-1}$ in an $\Omega=1$, $\Lambda=0$, $H_0=70~{\rm
km~s^{-1}~Mpc^{-1}}$ cosmology.

We model the frequency and time dependence of the afterglow luminosity,
$L_\nu(t)$, using the simplest unbeamed synchrotron model (e.g. Waxman
1997a, 1997b), with a luminosity per unit emitted frequency ${\tilde \nu}$
of
\begin{equation}
L_{\tilde \nu}(t) = L_{\nu_m}\left[\frac{{\tilde
\nu}}{\nu_m(t)}\right]^{-\alpha}\,,
\label{eq:lum}
\end{equation}
where $\nu_m(t)=8.8\times 10^{2}(1+z)^{1/2} (t/{\rm month})^{-3/2}\;{\rm
GHz}$, $t$ is time at the source frame, and $z$ is the source redshift.
The observed frequency $\nu$ is related to the emitted frequency ${\tilde
\nu}$ through $\nu={\tilde \nu}/(1+z)$. The spectral index $\alpha$ is
chosen to have the values $\alpha_1=1/3$ for $\nu\le\nu_m$ and
$\alpha_2=0.7$ for $\nu>\nu_m$, so as to match the temporal decay slope
observed for GRB 970228 (Galama et al.  1997; Fruchter et al.  1998b) and
GRB 970508 (Galama et al. 1998).

We consider a population of afterglow sources characterized by a total
comoving rate per unit volume, $R(z)$, and by a peak flux
$F_{\nu_m}(z,L_{\nu_m})=L_{\nu_m}(1+z)/4\pi D^2_{\rm L}(z)$, where $D_{\rm
L}(z)$ is the cosmology-dependent luminosity distance. For consistency with
the numbers derived by Wijers et al. (1997) we assume $\Omega=1$,
$\Lambda=0$, $h=0.7$, and $D_{\rm L}(z)=(2c/H_0)(1+z-\sqrt{1+z})$.  Note
that if the radio flux of GRB sources is strictly proportional to their
$\gamma$--ray flux, then our number count results are independent of the
choice of cosmological parameters to zeroth order; this follows from the
fact that Wijers et al. (1997) calibrate their results based on GRB number
count data.

We allow for a spread in the peak luminosity $L_{\nu_m}$ of the afterglows
by using a log-normal probability distribution (and thus minimizing the
number of free parameters),
\begin{equation}
P(L_{\nu_m})dL_{\nu_m} = \frac{1}{\sqrt{2\pi\sigma^2}}
\exp\left\{-\frac{[\ln(L_{\nu_m})-\ln(L_\star)]^2}{2\sigma^2}\right\}
{dL_{\nu_m}\over L_{\nu_m}}\;.
\label{eq:pl}
\end{equation}
The mean of this distribution equals $\langle
L_{\nu_m}\rangle=L_\star\exp\left(\sigma^2/2\right)$. We normalize its
value by matching the mean peak flux,
\begin{equation}
\langle F_{\nu_m}\rangle=\frac{\int_0^\infty dL_{\nu_m}P(L_{\nu_m})
\int_0^\infty dz [R(z)/(1+z)] F_{\nu_m}(z,L_{\nu_m})dV_c} {\int_0^\infty
dL_{\nu_m}P(L_{\nu_m}) \int_0^\infty dz [R(z)/(1+z)] dV_c}\;,
\label{Fave}
\end{equation}
to its observed value in radio afterglows.  The observed 8.44 GHz
lightcurves of GRB970508, GRB980329, and GRB980703 yield $\langle
F_{\nu_m}\rangle \approx 0.7$ mJy. For the above cosmology, this implies
$\langle L_{\nu_m}\rangle\approx 1.5\times 10^{31}\;h_{70}^{-2} $ erg ${\rm
sec}^{-1}\; {\rm Hz}^{-1}$.

The calculation of the afterglow number counts should also incorporate the
fact that not all GRBs have afterglows at radio frequencies. The fraction
$f_{\rm radio}$ of radio-loud GRBs is still uncertain, but the observation
that 4 out of $\sim 15$--17 well-localized GRBs (namely, GRB970508,
GRB980329, GRB980519 and GRB980703) were detected in the radio (D. Frail
1998, private communication), implies that $f_{\rm radio}\simeq 25\%$.
%Rosalba: please show results for f_radio=25% in the plots. Thanks.
Also, the afterglow radio flux might have a break in its power-law
evolution and decay more rapidly than predicted by equation~(\ref{eq:lum})
after a particular time (e.g., during the non-relativistic phase of the
shock hydrodynamics). We artificially introduce a free parameter, $t_{\rm
cutoff}$, after which we set the emitted radio flux to zero.  Current data
does not allow an empirical calibration of $t_{\rm cutoff}$ and so we leave
it as a free parameter. We assume that $t_{\rm cutoff}\ga 3$ months since a
simple power-law decline was observed in GRB970508 over 3 months (Waxman et
al.  1998).

The late phase of the radio emission, several months after a GRB event,
originates from semi-relativistic material with $\Gamma\sim 1$ and receives
only weak relativistic beaming. However, the $\gamma$-ray emission is
emitted by material with $\Gamma\ga 10^2$ and could be confined to a cone
with an angular diameter $\theta_{\rm b}\ll 1$. In this case, the rate of
radio afterglows would be enhanced by a factor $f_{\rm
b}^{-1}=4\pi/\pi(\theta_{\rm b}/2)^2=5.25\times 10^2(\theta_{\rm
b}/10^\circ)^{-2}$, relative to the unbeamed case.

Typical radio surveys do not provide continuous monitoring of the sky, but
rather a sequence of ``snapshots'' of each resolution element on the sky.
In observations where the minimum detectable flux of a source is $F_\nu$,
the total number of events across the sky that are brighter than $F_\nu$ at
observed frequency $\nu$ is given by
\begin{equation}
N(>F_\nu;\nu)= f_{\rm b}^{-1} f_{\rm radio} \int_0^\infty
dL_{\nu_m}P(L_{\nu_m})N'(>F_\nu;\nu,L_{\nu_m})
\label{eq:numb}
\end{equation}
where
\begin{equation}
N'(>F_\nu;\nu,L_{\nu_m})=\int_0^{z_{\rm lim}(L_{\nu_m})}\, R(z) \,
t_*(z,F_\nu,\nu,L_{\nu_m}){dV_c\over dz} dz\;.
\label{eq:nl}
\end{equation}
Here $dV_c=4(c/H_0)^3(1+z-\sqrt{1+z})^2(1+z)^{-7/2} d\Omega dz$~~is the
comoving volume element within a solid angle $d\Omega$ and redshift
interval $dz$, and $z_{\rm lim}(L_{\nu_m})$ is found from the algebraic
relation $F_{\nu_m}(z_{\rm lim},L_{\nu_m})=F_\nu$.  The time
$t_*(z,F_\nu,\nu)$ represents the duration over which an event at a
redshift $z$ is brighter than the limiting flux $F_\nu$ at the observed
frequency $\nu$,
\begin{equation}
t_*(z,F_\nu,\nu,L_{\nu_m})={\rm min}\left\{t_{\rm cutoff},\;
\left(\frac{(1+z)\nu}{\nu_{m_0}}\right)^{-2/3}
\left[\left(\frac{F_\nu}{F_{\nu_m}(z,L_{\nu_m})}\right)^{-\frac{2}{3\alpha_2}}-
\left(\frac{F_\nu}{F_{\nu_m}(z,L_{\nu_m})}\right)^{-\frac{2}{3\alpha_1}}
\right] \right\},
\label{eq:t*}
\end{equation}
where $\nu_{m_0}\equiv 8.8\times 10^2~(1+z)^{1/2}{\rm GHz~month^{3/2}}$.

\section{Constraints from Radio Surveys}

The VLA FIRST survey monitored 1550 degrees of sky with a sensitivity of 1
mJy at a frequency of 1.5 GHz (Helfand et al. 1996). The intervals between
the observations of each source ranged between 3 minutes and 3 weeks.
Unfortunately, the properties of this survey were not optimal for the
purpose of identifying GRB afterglows.  Here we consider a hypothetical
survey which is optimized for this purpose.  Such a survey would use a
higher frequency in order to avoid the suppression of the afterglow flux by
synchrotron self-absorption at the source (typically observed at $\la 5
[(1+z)/2]^{-1}$ GHz).  Moreover, since short-term variability could also be
caused by scintillations, the surveyed sources should be monitored at least
twice over an extended period of order a month or more.  After a few months
all the radio-loud afterglows would fade significantly and thus be easily
distinguishable from other variable source populations.  The variability
level of $\sim 0.1$--$1$ mJy sources can be determined accurately, given
the high sensitivity of the VLA.

Figure 1 shows the number of radio afterglows that are expected to be found
in an all-sky survey at 10 GHz, as a function of its flux threshold.  The
plotted number depends on a normalization factor involving the
$\gamma$--ray beaming angle $\theta_{\rm b}$ in units of $10^\circ$.  We
assume $f_{\rm radio}=25\%$ and show results for different values of
$\sigma$ while keeping $t_{\rm cutoff}$ fixed at 6 months. Figure 2 shows
the same for $\sigma \rightarrow 0$ but for different values of $t_{\rm
cutoff}$.  Typically, a 0.1 mJy survey similar to FIRST but at 10 GHz would
identify $\ga 2\times 10^4(\theta_{\rm b}/10^\circ)^{-2}$ afterglows across
the sky if the $\gamma$-ray emission from GRBs is collimated to within an
opening angle $\theta_{\rm b}$.

Lower limits on the value of $\theta_{\rm b}$ can already be placed by
existing radio surveys. The total number of radio sources observed on the
sky provides a robust upper limit on the number of GRB afterglows. This
limit can be improved by considering only those sources which are
unresolved and at the same time variable over a time scale of a month or
longer. Still tighter constraints can be obtained by considering only the
subset of all unidentified sources that after a sufficiently long time,
$\ga$ 6 months, fade away below the detection threshold.  From a deep VLA
survey, Windhorst et al. (1993) derived a maximum-likelyhood fit to the
source number counts at 8.44 GHz in the flux range of 14.5$\mu$Jy--1.5mJy.
By comparing a Westerbork and a VLA survey of the same field at 1.4 GHz,
Oort \& Windhorst (1985) found that about 10\% of the sources were variable
on a time scale of a year, and only $\sim 3\%$ were detected in one survey
but not in the other. Thus, only $\sim 3\%$ of the sources in this survey
qualify as candidates for GRB afterglows.  The sub-mJy sources were
typically found to be unresolved. Better angular resolution could in
principle limit further the number of afterglow candidates by resolving
other source populations.

Figure 3 shows the ratio between the predicted number of GRB afterglows
$N(>F_\nu)$ and the total observed number of radio sources $N_{\rm obs}$
(Windhorst et al. 1993) at 8.44 GHz.  We denote the fraction of variable
sources which fade away after a year by $f_{\rm fade}$, and the fraction of
pointlike (unresolved) sources by $f_{\rm point}$.  Based on the observed
abundance of unresolved variable sub-mJy sources at 1.4 GHz (Oort \&
Windhorst 1985), we set $f_{\rm fade}\times f_{\rm point}=3\%$.  The ratio
between the predicted afterglow counts and the observed radio counts peaks
around $0.4$--$0.6$ mJy. From the requirement that this ratio be smaller
than unity, we get the lower limit $\theta_{\rm b}\ga 6^\circ$. This limit
is based on the variability statistics of only $\sim 10^2$ sources (Oort \&
Windhorst 1985) and hence suffers from large statistical uncertainties
(min$\{\theta_{\rm b}\}\propto f_{\rm fade}^{-1/2}$).  Surveys with a
larger coverage of the sky are necessary in order to firm up the
statistical significance of this limit.

\section{Conclusions}

We have found that a $0.1$ mJy all-sky survey at 10 GHz should identify
$\sim 2\times 10^4 (\theta_{\rm b}/10^\circ)^{-2}$ GRB afterglows, and
could therefore place important constraints on the GRB beaming angle,
$\theta_{\rm b}$. In fact, the total number of $\ga0.1$ mJy sources on the
sky at 8.44 GHz (Windhorst et al. 1993) and the fraction of fading
unresolved sources at 1.4 GHz (Oort \& Windhorst 1985), already yield the
crude lower limit $\theta_{\rm b}\ga 6^\circ$. This result is only weakly
sensitive to the width of the afterglow luminosity function (see Fig. 1),
and is mainly uncertain due to the limited size of the sample of sub-mJy
sources that had been monitored for long-term variability (Oort \&
Windhorst 1985). Nevertheless, the derived constraint is interesting, given
the fact that the $\gamma$-ray emission in GRB sources originates from
material with a bulk Lorentz factor $\Gamma \ga 10^2$ and could have been
collimated to within an angle $\theta_{\rm b}\sim 1/\Gamma\la 0.6^\circ$,
narrower by more than an order of magnitude than our limit.  In comparison,
radio jets in quasars are collimated to within $\sim 15^\circ$ and possess
Lorentz factors $\Gamma\la 10$ (Begelman, Blandford, \& Rees 1984).

A future VLA survey, optimized to search for radio afterglows by monitoring
sub-mJy source variability at $\sim 10$ GHz over a timescale of several
weeks to several months, could improve the above upper limit on
$\theta_{\rm b}$ considerably.  Alternatively, such a search might identify
a new class of radio afterglow events which fade away after a few months
and have no GRB counterparts.

\acknowledgements 

We thank Dale Frail for useful discussions.  This work was supported in
part by the NASA grant NAG5-7039.

\begin{figure}[t]
\centerline{\epsfysize=5.7in\epsffile{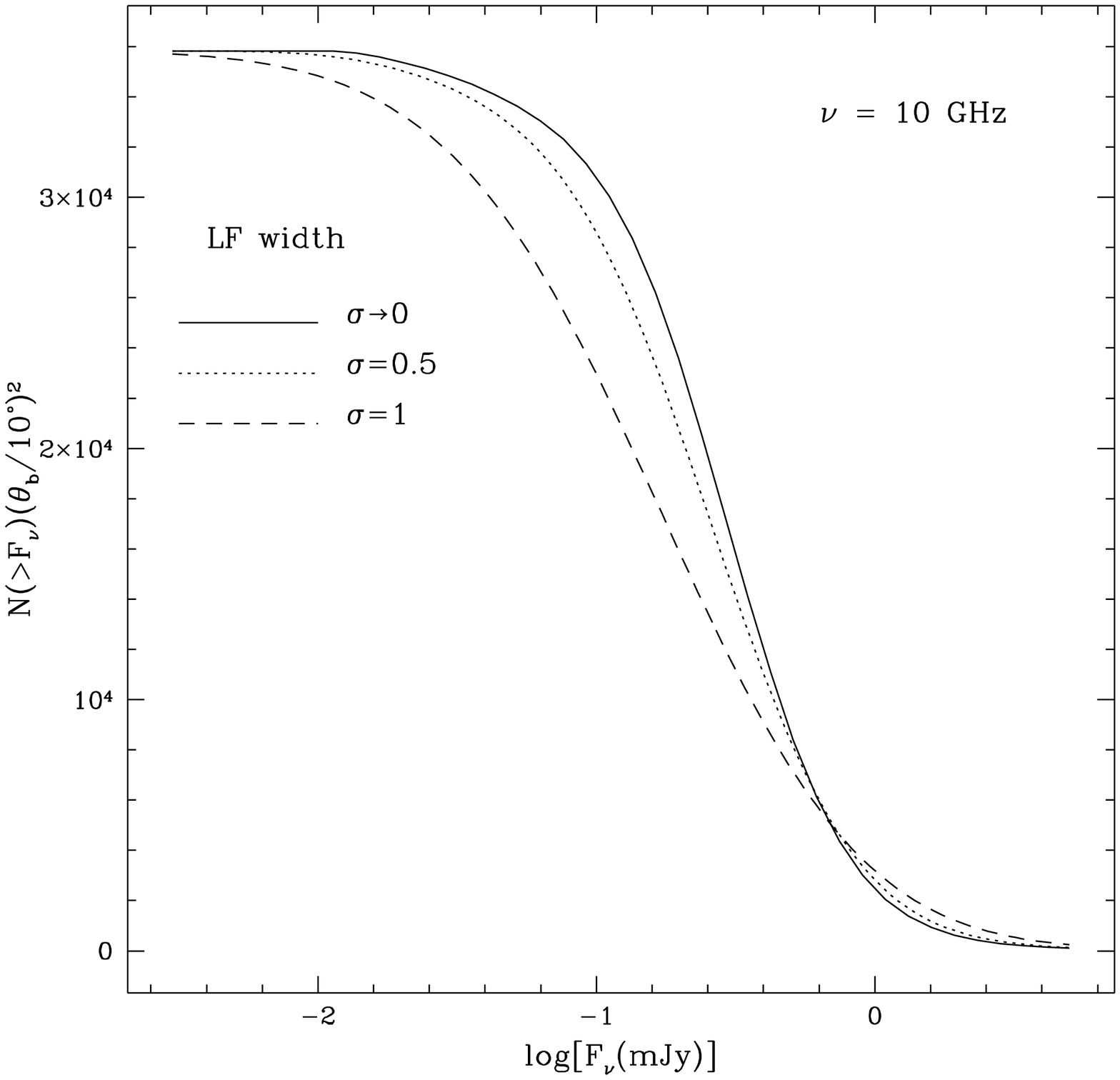}}
\caption{Number of radio-loud afterglows that would be detected in an
all-sky radio survey as a function of its flux detection threshold $F_\nu$
at 10 GHz.  The different lines correspond to different values of the width
of the GRB luminosity function, $\sigma$ [see Eq. (2)]. The fraction of the
radio-loud afterglows is assumed to be $f_{\rm radio}=25\%$ and the
afterglow emission is truncated after a time $t_{\rm cutoff}=6$ months in
all sources. The normalization factor of the number counts involves the
opening angle of the $\gamma$-ray emission, $\theta_{\rm b}$, in units of
$10^\circ$.}
\label{fig:1}
\end{figure}

\begin{figure}[t]
\centerline{\epsfysize=5.7in\epsffile{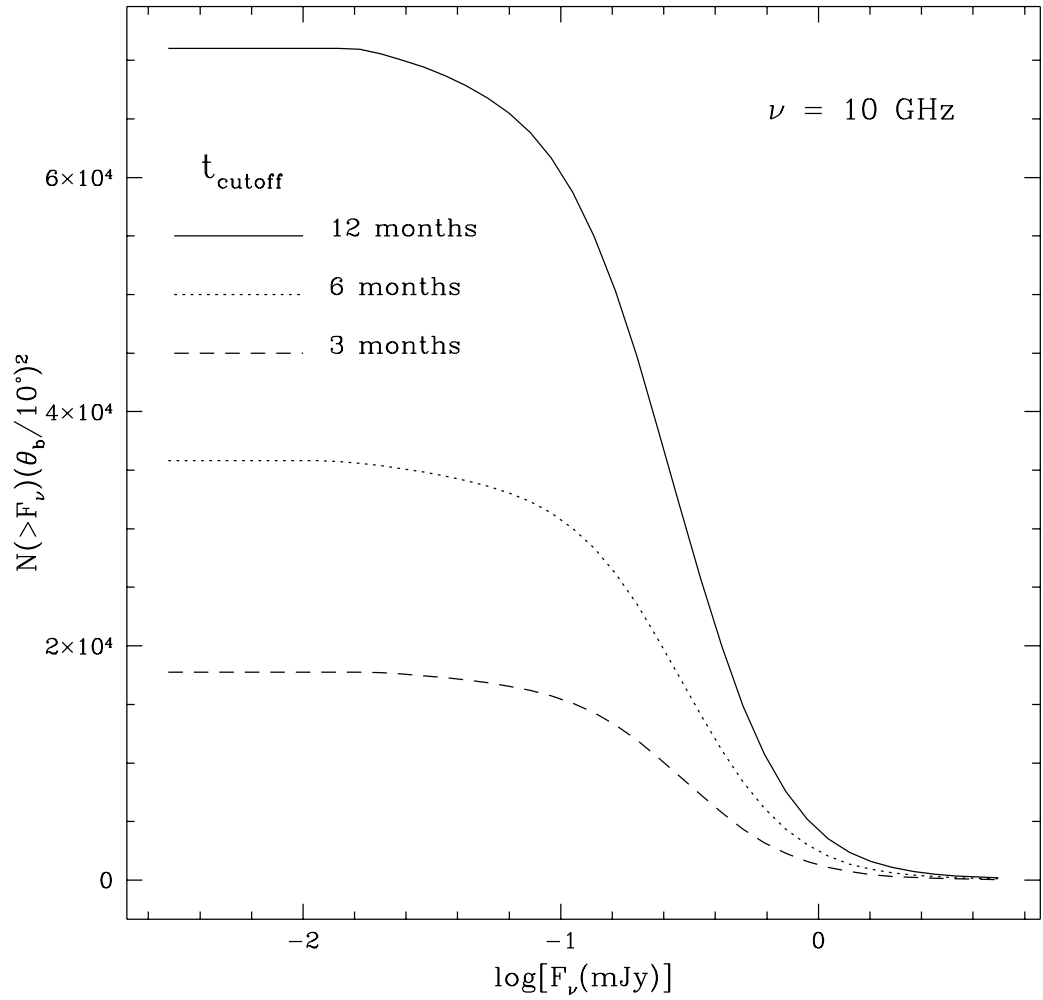}}
\caption{Same as in Figure 1, but with a fixed $\sigma \rightarrow 0$ and
different values of $t_{\rm cutoff}$.}
\label{fig:2} 
\end{figure} 

\begin{figure}[t]
\centerline{\epsfysize=5.7in\epsffile{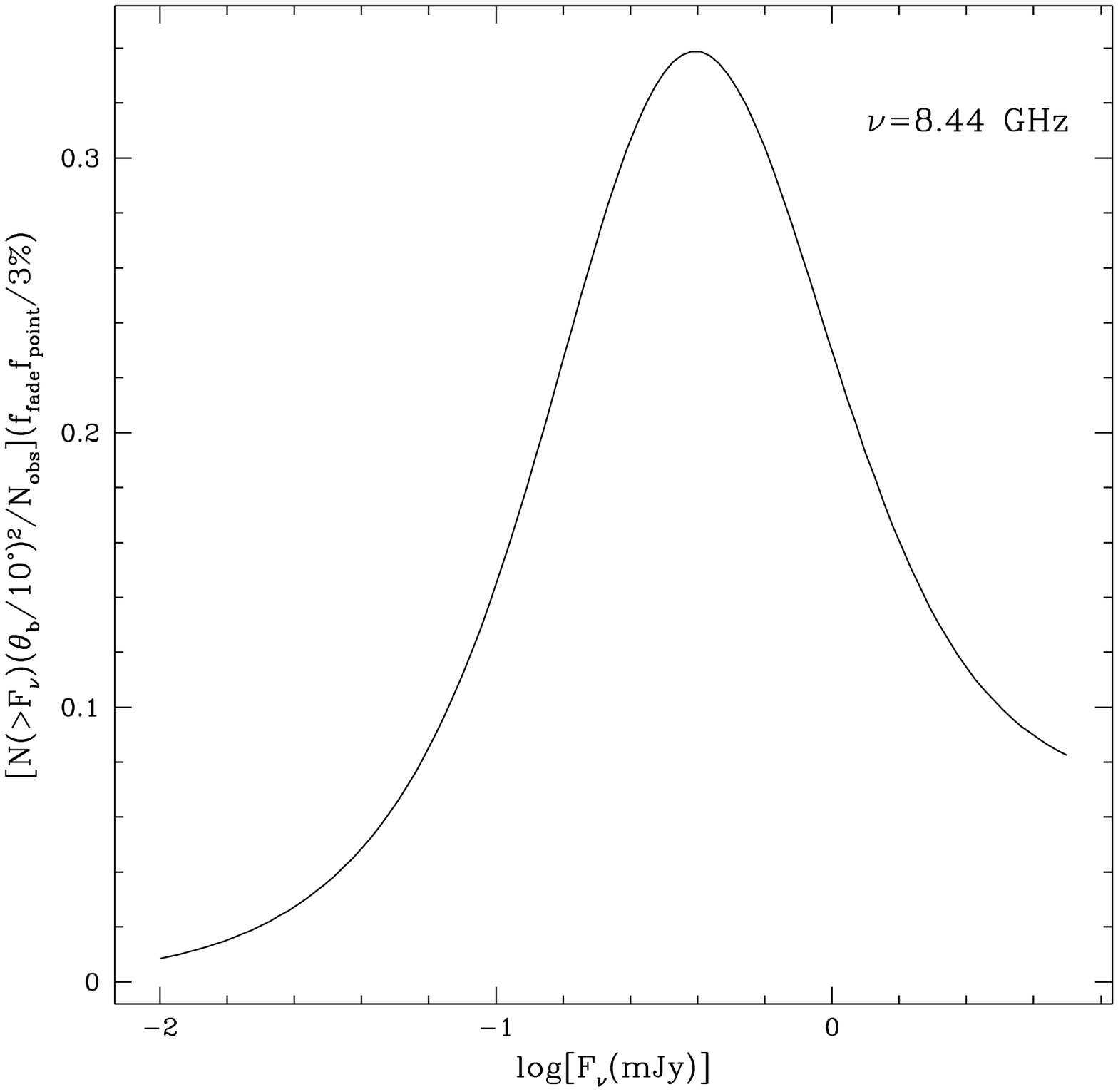}}
\caption{The ratio between the predicted cumulative number of GRB
afterglows $N(>F_\nu)$ and the observed cumulative number of radio sources
$N_{\rm obs}$ (Windhorst et al. 1993), as a function of the flux threshold
$F_{\rm \nu}$ at $8.44$ GHz.  The observed number is uncertain by $\pm
40\%$.  The fraction of unresolved fading sources in the observed
population is assumed to be $f_{\rm fade}f_{\rm point}=3\%$, based on 1.4
GHz data (Oort \& Windhorst 1985). Model parameters are chosen as in Figure
1, with $\sigma \rightarrow 0$.  The condition that the plotted ratio be
less than unity at any flux threshold sets a lower limit on the
$\gamma$-ray beaming angle $\theta_{\rm b}\ga 6^\circ$.}
\label{fig:3} 
\end{figure}


\begin{references}

\reference{}
Ali, S., Schaefer, R. K., Limon, M., \& Piccirillo, L. 1997, ApJ, 487, 114
\reference{}

\reference{} 
Begelman, M. C., Blandford, R. D., Rees, M. J. 1984,
Mod. Phys. Rev., 56.2, 255 

\reference{} Bond, H. E. 1997a, IAU Circ. No. 6654 

\reference{} Bond, H. E. 1997b, IAU Circ. No. 6665

\reference{} Briggs, M. S., et al. 1993, in Proc. of the Huntsville
Gamma-Ray Burst Workshop, ed. G. Fishman, J. Brainerd, \& K. Hurley (New
York: AIP), 44 

\reference{} Costa, E. et al. 1997, Nature, 387, 783
\reference{}
Dessenne, C. A. C. et al. 1996, MNRAS, 281, 977
\reference{} Djorgovski, S. G., Kulkarni, S. R., Bloom, J. S., Goodrich,
R., Frail, D. A., Piro, L., \& Palazzi, E. 1998, ApJL, submitted,
astro-ph/9808188

\reference{} Eichler, D., Livio, M., Piran, T., \& Schramm, D. N. 1989,
Nature, 340, 126 

\reference{} Fenimore, E. E., Epstein, R. L., \& Ho,
C. 1993, A\&AS, 97, 59 

\reference{} Frail, D. A. 1998, in Proceedings of Gamma Ray Bursts, 4th
Huntsville Symposium, AIP Conf. Proc. 428, eds, C. A. Meegan, R. D. Preece,
T.  M. Koshut, (AIP Press: New York), pp. 563--569

\reference{} Frail, D. A. et al. 1997a, ApJ, 483L, 91

\reference{} Frail, D. A., et al. 1997b, Nature, 389, 261 

\reference{} Frail, D. A., Taylor, G., \& Kulkarni, S. R. 1998a, GCN
Circ. No. 89 

\reference{} Frail, D. A., Halpern, J. P., Bloom, J. S., Kulkarni, S. R.,
\& Djorgovski, S. G. 1998b, GCN Circ. No. 128

\reference{} Fruchter, A. S. et al. 1998a, astro-ph/9807212 

\reference{} Fruchter, A. S., et al. 1998b, to appear in Proc. of the 4th
Huntsville Symposium on Gamma-Ray Bursts, eds. C. A.  Meegan, R. Preece, \&
T. Koshut, astro-ph/9801169


\reference{} Fryer,  C. L., Woosley, S. E., Herant, M., \& Davies, M. B.
1998, ApJ, submitted, astro-ph/9808094

\reference{} Galama, T. J., et al., 1997, to appear in Proc. of the 4th
Huntsville Symposium on Gamma-Ray Bursts, eds. C. A.  Meegan, R. Preece, \&
T. Koshut, astro-ph/9712322


\reference{} Galama, T. J., et al. 1998, ApJL, 500, 101 

\reference{} Helfand, D. J., Das, S. R., Becker, R. H., White, R. L.,
McMahon, R. G.  1996, in Blazar Continuum Variability, eds. H. R. Miller,
J. R. Webb, \& J. C. Noble (Astron. Soc. of the Pacific: San Francisco),
pp. 214--219

\reference{} Kulkarni, S. R. et al. 1998a, Nature, 393, 35 

\reference{} Livio, M., et al. 1997, ApJL, 489, L127

\reference{} Madau, P. 1996, in Star Formation Near and Far, AIP
Conf. Proc.  (New York: AIP), astro-ph/9612157 

\reference{} Madau, P. 1997, to appear in The Hubble Deep Field, ed.
M. Livio, S. M. Fall, \& P. Madau, STScI Symposium Series, astro-ph/9709147

\reference{} Madau, P., Ferguson, H. C., Dickinson, M. E., Giavalisco, M.,
Steidel, C.  C., \& Fruchter, A. 1996, MNRAS, 283, 1388 

\reference{} Madau, P., Pozzetti, L., \& Dickinson, M. E. 1998, ApJ, 498,
106 

\reference{} Meegan, C. A., et al. 1993, in Proc. of the Huntsville
Gamma-Ray Burst Workshop, ed. G. Fishman, J. Brainerd, \& K. Hurley (New
York: AIP), 3 

\reference{} Meszaros, P., \& Rees, M. J. 1997, 476, 232

\reference{} Meszaros, P., Rees, M. J., Wijers, R.A.M.J. 1998,
New Astronomy, submitted, astro-ph/9808106

\reference{} Metzger, R. et al. 1997, Nature, 387, 878

\reference{} Narayan, R., Paczy\'nski, B., \& Piran, T. 1992, ApJ, 395, L83

\reference{} Oort, M. J. \& Windhorst, R. A. 1985, A\&A, 145, 405

\reference{} Paczy\'nski, B. 1986, ApJ, 308, L43 

\reference{} Paczy\'nski, B. 1998, ApJ, 494, L45 

\reference{} Popham, R., Woosley, S. E. \& Fryer, C. 1998, astro-ph/9807028

\reference{} Rees, M. J. 1998, in Proc. of the 18th Texas symposium on
Relativistic Astrophysics and Cosmology, eds. A. V. Olinto, J. A. Frieman,
\& D. A. Schramm, (World-Scientific: Singapore), p. 34 ; astro-ph/9701162

\reference{} Rhoads, J. E. 1997, ApJ, 487, L1 

\reference{} Rhoads, J. E. 1998, to appear in Proc. of the 4th
Huntsville Symposium on Gamma-Ray Bursts, eds. C. A.  Meegan, R. Preece, \&
T. Koshut,  astro-ph/9712042

\reference{} Sokolov, V. V., et al. 1998, A \& A, 334, 177

\reference{} Taylor, G., Frail, D. A., \& Kulkarni, S. R. 1998, GCN
Circ. No. 40

\reference{} Usov, V. V. 1992, Nature, 357, 472 

\reference{} van Paradijs, J., et al. 1997, Nature, 386, 686 
\reference{}
Wall, J. V. 1994, Australian J. Phys., 47, 625
\reference{} Waxman, E. 1997a, ApJ, 485, L5 

\reference{} ---------------. 1997b, ApJ, 489, L33

\reference{} Waxman, E. Kulkarni, S. R. \& Frail, D. A. 1998,
ApJ, 497, 288 

\reference{} Wijers, R. A. M. J., Bloom, J. S., Bagla, J. S., \& Nataraja,
P. 1997, MNRAS, 294, L13 
\reference{}
Windhorst, R. A., Fomalont, E. B., Partridge, R. B. \& Lowenthal, J. D.
1993, 405, 498
\reference{} Woods, E., \& Loeb, A. 1995, ApJ, 453, 583 

\reference{} -----------------------------. 1998, ApJ, in press;
astro-ph/9803249 \reference{} Woosley, S. E. 1993, ApJ, 405, 273

\reference{} Woosley, S. E. 1998, ApJ, 405, 273

\end{references}
\end{document}